\documentclass[reprint, amsmath,amssymb, aps, superscriptaddress, floatfix]{revtex4-1}
\usepackage{graphicx}
\usepackage{dcolumn}
\usepackage{bm}

\usepackage{amsbsy,amssymb,latexsym,amsfonts,amsmath}
\usepackage{graphicx,color}
\allowdisplaybreaks

\begin{document}

\preprint{APS/123-QED}
\title{RG scaling relations at chiral phase transition in two-flavor QCD}

\author{K.-I. Ishikawa}
\affiliation{Graduate School of Science, Hiroshima University,Higashi-Hiroshima, Hiroshima 739-8526, Japan}

\author{Y. Iwasaki}
\affiliation{Center for Computational Sciences, University of Tsukuba,Tsukuba, Ibaraki 305-8577, Japan}

\author{Yu Nakayama}
\affiliation{Kavli Institute for the Physics and Mathematics of the Universe (WPI), Todai Institutes for Advanced Study,  Kashiwa, Chiba 277-8583, Japan}
\affiliation{Department of Physics, Rikkyo University, Toshima, Tokyo 177-8501, Japan}

\author{T. Yoshie}
\affiliation{Center for Computational Sciences, University of Tsukuba,Tsukuba, Ibaraki 305-8577, Japan}

\date{\today}

\begin{abstract}

We investigate the nature of the chiral phase transition in the massless two-flavor QCD using the renormalization group improved gauge action and the Wilson quark action on $32^3\times 16$, $24^3\times 12$, and $16^3\times 8$ lattices. Based on the renormalization group equation, we derive the scaling relation for the effective masses of mesons at the chiral phase transition point. If the chiral phase transition is 
second order, the effective masses as a function of the rescaled time/space do not depend on the lattice size
 and show the universal behavior. We find that our numerical simulations on the three  sizes of lattices are excellently on the scaling curves, which is consistent with the second order phase transition.
\end{abstract}

\maketitle

\section{introduction}
\label{introduction}
The properties of quarks and gluons at high temperatures and the nature of the transition from the quark-gluon state to the hadronic state
are key ingredients for understanding the
evolution of the Universe. Yet, even the order of the chiral phase transition is still under debate. To make the discussions more transparent, let us consider the idealistic situation in which we have 
 two massless quarks in the SU(3) gauge theory, and ask the question whether the finite temperature phase transition of the chiral
symmetry is first order or second order.
 \cite{Pisarski:1983ms} - \cite{Nakayama:2014sba}
 
Lattice QCD is probably the most reliable constructive formulation for the investigation of  non-perturbative properties of quarks and gluons at the chiral phase transition point (see e.g. \cite{review1} for a review).
In this article we investigate the chiral phase transition in the massless two-flavor QCD, using the renormalization group (RG) improved gauge action and the Wilson quark action  with two degenerate quarks on $32^3\times 16$, $24^3\times 12$, and $16^3\times 8$ lattices.
The simulation itself may be a standard one, but we are going to employ a new method based on the renormalization group to investigate the long distance physics of the chiral phase transition.

The key idea is to test the RG scaling relation of the meson propagator. Based on Wilson's idea of RG, we derive the RG equation for the effective  masses of mesons. 
The RG equation we derive contains only two relevant (including marginal relevant) operators (i.e. the gauge coupling and the quark mass), and the RG equation has an interesting scaling solution at the chiral phase transition if it is second order. Alternatively, when the chiral phase transition is first order, then we do not expect the existence of such a solution,
or even the RG equation as it is may not hold because of the emergence of other relevant operators.

We show that effective masses calculated on the three lattices for the pseudo scalar (PS) mesons and vector (V) mesons 
agree with the predicted scaling relations in an excellent manner. 
While we cannot exclude the possibility that the behavior 
changes at the larger lattice size or in the thermodynamic limit, 
we argue that the numerically verified scaling behavior supports the claim that the chiral phase transition in the massless two-flavor QCD is second order.

The organization of the paper is as follows.
After describing our setup in section 2, we revisit RG equations in section 3 and derive RG scaling relations in section 4.
 After giving the job parameters in section 5, we identify the chiral phase transition points in section 6.
 We show our numerical results with the verification of the scaling relation in section 7.
The characteristic of the universal scaling curves is discussed in section 8.
In section 9, the chiral phase transition temperature is estimated.
  Finally summary and  discussion are given in section.10.

\section{Action and Observables}
\label{action}
We define continuous gauge theories as the continuum limit of lattice gauge theories,
defined on the Euclidean lattice of the size $N_x=N_y=N_z=N_s$ and $N_t$.  We impose an anti-periodic boundary condition in the temporal direction for fermion fields and periodic boundary conditions otherwise.

The continuum limit of QCD is taken at or toward a UV critical point (in this article we consider only the case where the UV fixed point is $g_0=0$ and $m_0=0$), 
with the lattice spacing $a \rightarrow 0$ together with $N_s \rightarrow \infty$ and $N_t \rightarrow \infty,$ changing the coupling $g(a)$ in a such way that physical quantities with physical dimension such  as hadron masses $m_{H} /a$  fixed.  
We may use the aspect ratio $r=N_s/N_t$ and $N=N_t$ instead of $N_s$ and $N_t.$
In this article, we keep the aspect ratio $r$ fixed when we change the lattice size.  The necessary change of the coupling constant is governed by the renormalization group equation.

When $N_s$ and $N_t$  are finite, the formulation of finite temperature QCD on a lattice is equivalent to a Euclidean path integral defined on a discrete three dimensional lattice cite with a transfer matrix for a discrete time. 
However,  at finite temperature we have to take the thermodynamic limit $r  \rightarrow \infty$
for interpretation of physical quantities at finite temperature.

Although our methodology can be applied to any gauge theories with fermions in arbitrary representation,
we focus on $SU(3)$ gauge theories with degenerate $N_f=2$ fundamental fermions (``quarks") in this article.
We employ the Wilson quark action and the RG improved gauge action~\cite{RG-improved}.
The theory is defined by two parameters; the bare coupling constant $g_0$ and the bare degenerate quark mass $m_0$ at ultraviolet (UV) cutoff.
We also use, instead of $g_0$ and $m_0$, 
$\beta={6}/{g_0^2}$
and the hopping parameter
$K= 1/2(m_0a+4)$. 

We measure the plaquette and the Polyakov loop in each space-time direction, the mass of hadrons such as the pseudo-scalar meson mass $m_{\text{PS}}$.
The quark mass $m_q$ is defined~\cite{Bochicchio:1985xa},\cite{Itoh:1986gy}.as the large $t$ value of $m_q(t)$ obtained through Ward-Takahashi identities by the ratio of thermal propagators=
\begin{align}
m_q &= \lim_{t  \to N_t/2}m_q(t) \cr
  &= \lim_{t  \to N_t/2} \frac {\sum_{x} \langle \nabla_4 A_4(x,t) P(0) \rangle }{ 2\, \sum_{x} \langle P(x,t) P(0) \rangle}
\label{quark mass}
\end{align}
where $P(x,t)$ is the pseudo-scalar density and $A_4(x,t)$ the fourth component of the
local axial vector current, renormalization constants being suppressed.
This is also the same definition that we use in the zero-temperature lattice QCD.
The quark mass $m_q$ thus defined does not depend on whether the system is confining or deconfining, and
depends on only $\beta$ and $K$ up to order $1/N_s$ and $1/N_t$  corrections.

The propagator is defined by
\begin{align}
G_t(t) = \sum_{x,y,z} \langle \bar{\psi}\gamma_H\psi(x,y,z,t) \bar{\psi}\gamma_H\psi(0,0,0,0) 
\end{align}
for the temporal one and
\begin{align}
G_{s}(x) = \sum_{t,y,z} \langle \bar{\psi}\gamma_H\psi(x,y,z,t) \bar{\psi}\gamma_H\psi(0,0,0,0) 
\end{align}
for the spatial one. $H$ here stands for the channel, and in this article, we focus on the pseudo-scalar $(H=PS)$ and the vector channel $(H=V)$, and we sometimes omit the subscript $H$ in the following.

We define the effective mass $m(t)$ (or similarly $m(x)$) through  
\begin{equation}\frac{\cosh(m(t)(t-N_t/2))}{\cosh(m(t)(t+1-N_t/2))}=\frac{G(t)}{G(t+1)}.\label{effective mass}\end{equation}
which reduces to
\begin{equation}m(t) = \ln \frac{G(t)}{G(t+1)}.\label{simple effective mass}\end{equation}
 when boundary effects can be neglected.
 In the case of exponential-type decay the effective mass approaches a constant value 
in the large $t$ regime, which we call a plateau.

In our earlier studies Refs.~\cite{Ishikawa:2013wf} - \cite{Ishikawa:2013tua}, we argued that there is a region where the propagator exhibits a power-law corrected Yukawa type decay when the quark mass is small
on the  $16^3\times 64$ lattice (with aspect ratio $r=N_s/N_t=1/4$).
We observed the similar behavior for the aspect ratio $r=2$ which is larger than unity in  \cite{Ishikawa:2015nox}. In this article, we mainly discuss the $r=2$ case with the thermodynamic limit in mind. 

\section{RG equations}
\label{continuum limit}
Let us recall the RG equations on a finite lattice, which represent Wilson's basic idea~\cite{Wilson:1973jj}.
In the vicinity of a critical point where the correlation length diverges, short-range fluctuations can be integrated out while keeping the long distance behavior. 
When we change the size of the lattice, the action must be changed to keep the physics intact, but the key observation by Wilson is that near a critical point, 
it is sufficient to take into account relevant  (including marginal relevant) operators.

Now let us consider the case when the chiral phase transition is second order. 
The RG equation in the vicinity of the critical point is given 
~\cite{DelDebbio:2010ze} - \cite{DelDebbio:2010hu}
\begin{eqnarray}G_t(n_t; g, m_q, N, \mu)_H &=& \nonumber \\
{\left(\frac{N'}{N}\right)}^{-2\gamma} &G_t& (n_t'; g{'}, m'_q, N{'},  \mu{'} )_H \label{t-RG}\end{eqnarray}
for the temporal propagator
and 
\begin{eqnarray}G_s(n_s; g, m_q, N, \mu)_H &=& \nonumber \\
{\left(\frac{N'}{N}\right)}^{-2\gamma} &G_s& (n_s{'}; g{'}, m'_q, N{'},  \mu{'} )_H\label{s-RG} \end{eqnarray}
for the spatial propagator.
 The suffix $s$ and $t$ represents spatial and temporal, respectively. In the following we often suppress the suffix except when it is better to specify which we refer to.

Here $n=(n_x, n_y, n_z, n_t)$, and $\mu{'}= \mu/s $ and  $N_s{'}=N_s/s,$ $N_t{'}=N_t/s,$ and $n{'}=n/s$ with $s$ being the change of the scale under the renormalization.
The UV renormalization scale $\mu$ in lattice theories is set by the inverse lattice spacing $a^{-1}.$  
Note $N_s a=L_s$ and $N_t a =L_t$ are kept constant.
The relation between $g'$ and $g$ and $m'_q$ and $m_q$ are determined by the RG beta function $\mathcal{B}$ and the mass anomalous dimension $\gamma$: 
\begin{equation}\frac{d g}{d\log s} = \mathcal{B}\end{equation} and
\begin{equation}\frac{d m_q}{d\log s} = \gamma m_q.\end{equation} 

 To make the expression simpler, it is assumed that $\gamma$ does not change very much as the ratio $N'/N$ changes, which is always valid when we are at the RG fixed point, or when $N'/N \sim 1$ even if we are not at the fixed point. With these caveats in mind, we are going to verify the RG equation numerically.

 As a technical remark, we note that the RG equations (\ref{t-RG}) and (\ref{s-RG}) for the temporal propagator and for the spatial propagator define the same RG beta function and the mass anomalous dimension, since both are derived from the RG equation for the common two-point functions. However, the RG beta function $\mathcal{B}$ and the anomalous dimension $\gamma$ may depend on the aspect ratio $r$ and in addition the latter takes a different value for a different channel.

\section{RG Scaling relations}
Our goal is to test the RG equation to study a long distance behavior of the meson propagator.
For this purpose, let us rewrite the RG equation in a more convenient way.
We first define the scaled effective mass by
\begin{equation}\mathfrak{m}(n_t; g, m_q, N) = N \ln \frac{G(n_t; g, m_q, N)}{G(n_t+1; g, m_q, N)},\label{scaled effective mass}\end{equation}
suppressing $\mu$.
It reduces  in the continuum limit $N \rightarrow \infty$ to the form
\begin{equation}\mathfrak{m}(\tau; g, m_q, N)= - \partial_\tau \ln G(\tau; g, m_q, N).\label{effective mass relation}\end{equation}
Here $\tau =n_t/N_t$. 
The variable $n_t$ takes $0, 1, 2, \cdots, N_t-1$ so that the scaled time always lie in the range $0 \leq \tau \leq 1$.  

For the scaled effective mass, the RG equation takes the form
\begin{align}
\mathfrak{m}(\tau; g, m_q, N) =\mathfrak{m}(\tau; g', m'_q, N')
\end{align}
 
\subsection{Solution at an IR fixed point}

The RG equations~(\ref{t-RG}) and (\ref{s-RG}) derived at the chiral phase transition point are also applied to the IR fixed point of conformal QCD at zero temperature. Let us consider when the theory is in the conformal window  ($N_f*<N_f<16.5$)  at zero temperature.
 Suppose we are at an IR fixed point, then we have the fixed parameters  $g' = g = g^*$ and $m_q' = m_q= 0$ so that $\mathcal{B}=0$ and $\gamma = \gamma^*$ in the RG equation \label{scaling at IR}
 Eq.(\ref{t-RG}). Hence,  the propagator can be expressed by a  simplified notation suppressing $g=g^*, m_q=0,$
 \begin{equation} \tilde{G}(\tau,N)= G(\tau; g^*, m_q=0, N).\end{equation}
 and the RG relation eq.(\ref{t-RG}) reduces to
 \begin{equation}\tilde{G}(\tau; N)={\left(\frac{N{'}}{N}\right)}^{-2\gamma^*} \tilde{G}(\tau; N{'})\label{simple RG} \ . \end{equation}

 Combining Eqs.(\ref{effective mass relation}) and (\ref{simple RG}),
 we derive the RG scaling relations for effective masses
 \begin{equation} \mathfrak{m}(\tau, N)=\mathfrak{m}(\tau, N{'})\label{RG scaling}\end{equation}
 at the IR fixed point. 
 It means that the scaled effective mass does not depend on $N$ as a function of $\tau$.
 Therefore, the agreement of the scaled effective mass as a function of $\tau$ is a stringent test of the fixed point.
 We applied  in Ref.\cite{Ishikawa:2015iwa}  the scaling relation (\ref{RG scaling}) in order to identify the IR fixed point for $N_f= 16, 12,  8,  7$ in $SU(3)$ gauge theories. 

\subsection{Solution at chiral phase transition}
Let us assume that we are at a chiral phase transition point.
The chiral phase transition point depends on the lattice size $g(N)$, where the precise criterion to call the chiral phase transition point on a finite lattice will be discussed in section ~\ref{estimate}.
Using the simplified expression
\begin{equation} \check{G}(\tau; g(N), N)= G(t; g(N), m_q=0, N)\label{non_sim_equality of scaled effective mass_2}.\end{equation}
Eq. (\ref{t-RG})
reduces to 
\begin{equation}\check{G}(\tau; g(N), N)={\left(\frac{N{'}}{N}\right)}^{-2\gamma} \check{G}(\tau; g(N{'}),  N{'})\label{mass dimension}. \end{equation}

Correspondingly, the scaled effective mass becomes
\begin{equation}\mathfrak{\check{m}}(\tau, g(N) ,N)= - \partial_\tau \ln \check{G}(\tau, g(N), N)
\end{equation}
as in the previous subsection, and we obtain the RG equation 
 \begin{equation}\mathfrak{\check{m}}_t(\tau, g(N), N)=\mathfrak{\check{m}}_t(\tau, g(N{'}), N{'})\label{RG for t-mass}\end{equation}
and  \begin{equation}\mathfrak{\check{m}}_s(\tau, g(N), N)=\mathfrak{\check{m}}_s(\tau, g(N{'}),  N{'}).\label{RG for s-mass}\end{equation}
Eqs. (\ref{RG for t-mass}) and (\ref{RG for s-mass}) are key scaling relations of this article,
which are valid when the chiral phase transition is second order.
Since the number of the parameters is reduced by setting $g= g(N)$, one may solve the RG equation as
\begin{equation}
\mathfrak{\check{m}_{s}}(\tau, g(N), N)=\mathfrak{F_{s}(\tau)} , \label{universal}
\end{equation} 
where $\mathfrak{F}_{s}(\tau)$ is renormalization invariant.

The distinct point of Eqs.
(\ref{RG for t-mass}) and (\ref{RG for s-mass})
from Eq.(\ref{RG scaling}) is  that the coupling $g$ is fixed in the former, while $g$ moves according to the RG beta function in the latter.
Nevertheless, the both \eqref{RG scaling} and \eqref{universal} show the universal scaling behavior irrespective of $N$. 
Furthermore, while the RG equation does not say anything about the shape of the scaling curve $ \mathfrak{F_{s}(\tau)}$, we will see
the numerical result for scaled effective masses  $ \mathfrak{F_{s}(\tau)}$ at the chiral phase transition 
shows the very similar behavior to the one at the zero temperature in the conformal QCD. 

When the RG equation is evaluated in the vicinity of the UV fixed point $g_0=0$ and $m_0=0$,
the quark mass term is relevant and the gauge coupling is marginal.
Along the RG trajectory from the UV fixed point to the IR critical point, 
the beta function possesses a zero at a certain value $g=g^*$ in the conformal QCD, at which the quark mass term is relevant and the gauge coupling is irrelevant.
In the case of the chiral phase transition, it does not possess a zero, that is,  the beta function is negative 
along the RG trajectory. The gauge coupling constant (or temperature) is relevant, and we have to tune it to obtain the criticality.
More generally, we may consider an infinite parameter space of possible interactions $H= \sum g_i O_i$ and we may define the beta function $\beta_i(\mu)$ for each coupling $g_i.$
By RG transformation the effective Hamiltonian possess many possible interactions like a four-fermion interaction.
These operators are irrelevant operators at the UV fixed point and retain  irrelevant at the chiral phase transition point.
We have implicitly assumed that the number of relevant operators does not increase along the RG trajectory, which was at the core of the derivation of our RG equation.

In contrast, let us see what would happen if the chiral phase transition
were the first order phase transition. Then, typically,
we would expect an extra relevant operator that
will induce the chiral symmetry breaking. 
The search for such an operator in the effective Landau-Ginzburg model
was a key diagnosis for determining the order of phase transition in the 
study of Pisarski and Wilczek \cite{Pisarski:1983ms}.
Accordingly, if this scenario is the
mechanism of the first order phase transition, the assumption of our RG scaling relation is invalidated, and
there is no generic reason to believe that the scaling relation
that we proposed is observed.

\section{Job parameters}
We perform simulations with two degenerate quarks 
on $32^3 \times 16,$ $24^3 \times 12$  and $16^3 \times 8$ lattices  to investigate the scaling of the effective masses of mesons. 
The algorithm we employ is the blocked HMC algorithm \cite{Hayakawa:2010gm}.
We choose the run-parameters in such a way that the acceptance of the HMC Metropolis test is about $70\%\sim 90\%.$
The statistics are 1,000 MD trajectories for thermalization and $1000\sim5000$ MD trajectories for the measurement.
We estimate the errors by the jack-knife method
with a bin size corresponding to 100 HMC trajectories (see Table \ref{job parameter}).

\section{Identification of chiral phase transition points}
\label{estimate}
The first task in our numerical simulations is to determine the chiral phase transition point on the parameter space $K$ and $\beta$. For this purpose, let us discuss the nature of the chiral phase transition and its realization on a finite lattice.
At high temperature the chiral symmetry is restored and the order parameter (i.e. chiral condensate) vanishes:  $\,\langle\bar{ \psi}\psi \rangle = 0,$ 
while below the chiral phase transition temperature $\langle\bar{ \psi}\psi \rangle \neq 0$.
In the Wilson fermion formalism a properly subtracted $\langle\bar{ \psi} \psi \rangle$  can be defined via an
axial Ward-Takahashi identity~\cite{Bochicchio:1985xa} 
\begin{equation}\langle\bar{ \psi} \psi \rangle_{\text{sub}}=2\, m_q\, a\, (2K)^2 \sum_{t} G_{\text{PS}}(0, t),   \end{equation}
where $G_{\text{PS}}(0, t)$ is the propagator of the pseudo-scalar meson. 
 
 A direct approach to identify the (would-be) chiral phase transition points  in lattice simulations is  to  
 check whether the oder parameter of the chiral symmetry vanishes or does not. We should note, however, that there is no real phase transition on a finite lattice in this strict sense. The long-time average of order parameter is always zero in finite systems, and if we try to practically measure it over a finite time average, the change of the order parameter around the (would-be) phase transition will be rounded. 
 
 Rather than directly measuring the chiral condensate, we take the following alternative approach. Let us recall that the Banks-Casher relation~\cite{Banks:1979yr}\cite{DelDebbio:2010ze} states that the critical exponent 
 $\eta$ 
\begin{equation}  \lim_{m \to 0}  \langle  \bar{\psi} \psi(m) \rangle \simeq  m^\eta \end{equation}
is identical with the critical  exponent of eigenvalue density of the massless Wilson-Dirac operator.
\begin{equation}  \lim_{\lambda \to 0}\rho(\lambda) \simeq \lambda^\eta \end{equation}
Based on this fact, we proposed the ``on Kc method"~\cite{Iwasaki:1996zt}, which monitors the number of iteration of CG inversion" to determine the chiral phase transition points.
In the chiral symmetry broken phase, $\eta=0$ and it implies that the zero eigenvalues accumulate.
Hence it is impossible to invert the Dirac-Wilson operator in the chiral symmetry broken phase, while in the symmetric phase there are no theoretical obstruction for the inversion albeit many iterations may be needed in numerical methods.
We, therefore, proposed to identify the chiral phase transition as the point where we cannot invert the Dirac-Wilson operator to continue the numerical simulation.
 
Let us summarize our strategy in our setup. We first determine the hopping parameter $K_c$ for the massless quark from $\beta=2.3$ to $3.0$~(Table \ref{job parameter})
Then we estimate the chiral phase transition points on the lattices of the three sizes: $32^3\times 16$,  $24^3\times 12$ and
$16^3\times 8$ by using this ``on Kc method",

As mentioned above, on a finite lattice the phase transition is rounded and the change of the number  of iterations is not as drastic as it would be in the thermodynamic limit. Therefore there is certain arbitrariness in the definition of the phase transition on a finite lattice.
We adopt the following as our definition  of the (would-be) chiral phase transition point.
We begin with simulations on the massless lines of the $16^3\times 8,$ $24^3\times 24$
    and $32^3 \times 16$  lattices at some $\beta$ (we take $\beta=3.0$) which is apparently in the chiral symmetric phase and decrease $\beta$ by a step $\Delta \beta =- 0.1$.
As $\beta$ decreases, the number of the iterations for  the inversion of the Wilson-Dirac operator in molecular dynamics steps increases and the acceptance ratio decreases. 
Finally at some points we cannot continue the simulations without decreasing the molecular steps.

We identify such $\beta$'s as the chiral phase transition points. 
Our simulation shows that these three points are:
\begin{itemize}
\item
$ \beta_{*}\simeq 2.8; K_{*}=0.1455$ on the $32^3\times16$ lattice;
\item
$ \beta_{*}\simeq 2.6; K_{*}=0.1480$ on the $24^3 \times 12 $ lattice;
\item
$ \beta_{*}\simeq 2.3; K_{*}=0.1547$ on the $16^3 \times 8 $ lattice.
\end{itemize}

To double-check our criterion, we have also computed the chiral condensates on the $K_c$ in the chiral symmetric phase as given in Table 1:
They are of order $O(10^{-2})$ in lattice unit.
On the other hand, in the confining phase we are unable to compute the order parameter directly at the massless point. For reference
we quote some of them at small (not very small) quark mass: 
At $\beta =2.7$, $K=0145$ on the $32^3\times16$ lattice, where $m_q=0.049(1)$, we have $\,\langle\bar{ \psi}\psi \rangle=0.161(1)$.
At $\beta=2,5$, $K=0.149$ on the $24^3\times 12$ lattice, where $m_q=0.031(1)$, we have $\,\langle\bar{ \psi}\psi \rangle=0.103(1)$.
At $\beta=2.2$, $K=0.155$ on the $8^3\times16$ lattice, where $m_q=0.068(1),$ we have $\,\langle\bar{ \psi}\psi \rangle=0.296(5).$
They are of order $O(10^{-1})$ and are larger than those in the symmetric phase  by a factor of ten.
This is consistent with our identification of the (would-be) chiral phase transition points.

Now we introduce a discrete beta function $ \mathfrak{\hat{B}}(N_1, N_2)$ defined by
\begin{equation}
6/g_1^2 -6/g_2^2 = \mathfrak{\hat{B}}(N_1, N_2) \ln{(\mu_1 /\mu_2)}\label{new_beta}
\end{equation}
which relates the critical values $\beta^{*}$ to each other.
Note $N_i=L/a_i$ and $\mu_i=1/a_i.$ Therefore $\ln{(\mu_1/\mu_2)}=\ln{(N_1/N_2)}.$
This definition may be more convenient because the one-loop result shows that $\mathfrak{\hat{B}}$ is just a constant independent of $g$. From our results on the chiral phase transition point, we obtain 
\begin{equation}\mathfrak{\\\hat{B}(32, 16})=0.72\end{equation}
\begin{equation}\mathfrak{\\\hat{B}(32, 24})=0.69\end{equation}
\begin{equation}\mathfrak{\\\hat{B}(24, 16})=0.74\end{equation}
Thus the beta function  $\mathfrak{\hat{B}}(N_1, N_2)$ is roughly  a constant $0.7.$
It means the three equations are mutually consistent.
The sum of the left side of the three equations (\ref{new_beta}) vanishes and therefore the sum of the right sides also should vanish.
If we can regard $\mathfrak{\hat{B}}$ is a constant around here, the consistency is automatically satisfied \begin{equation}
\mathfrak{\hat{B}}( \ln{(\mu_1 /\mu_2)}+\ln{(\mu_2 /\mu_3)}+\ln{(\mu_3 /\mu_1)})=0 
\end{equation} 
Converting the result to the standard one,
$$\mathfrak{B}=-\frac{g^3}{12} \mathfrak{\hat{B}}$$
we obtain the beta function $\mathfrak{B}=-0.183$ at $\beta=2.8$ and $-0.246$ at $\beta=2.3$. We curiously note that this number is close to the one-loop beta function of $\mathcal{B} = -\frac{33-2N_f}{48\pi^2} g^3$.

\section{Numerical results for RG scaling}
\subsection{spatial effective masses}
\label{spatial mass}

\begin{figure*}
\includegraphics[width=7.5cm]{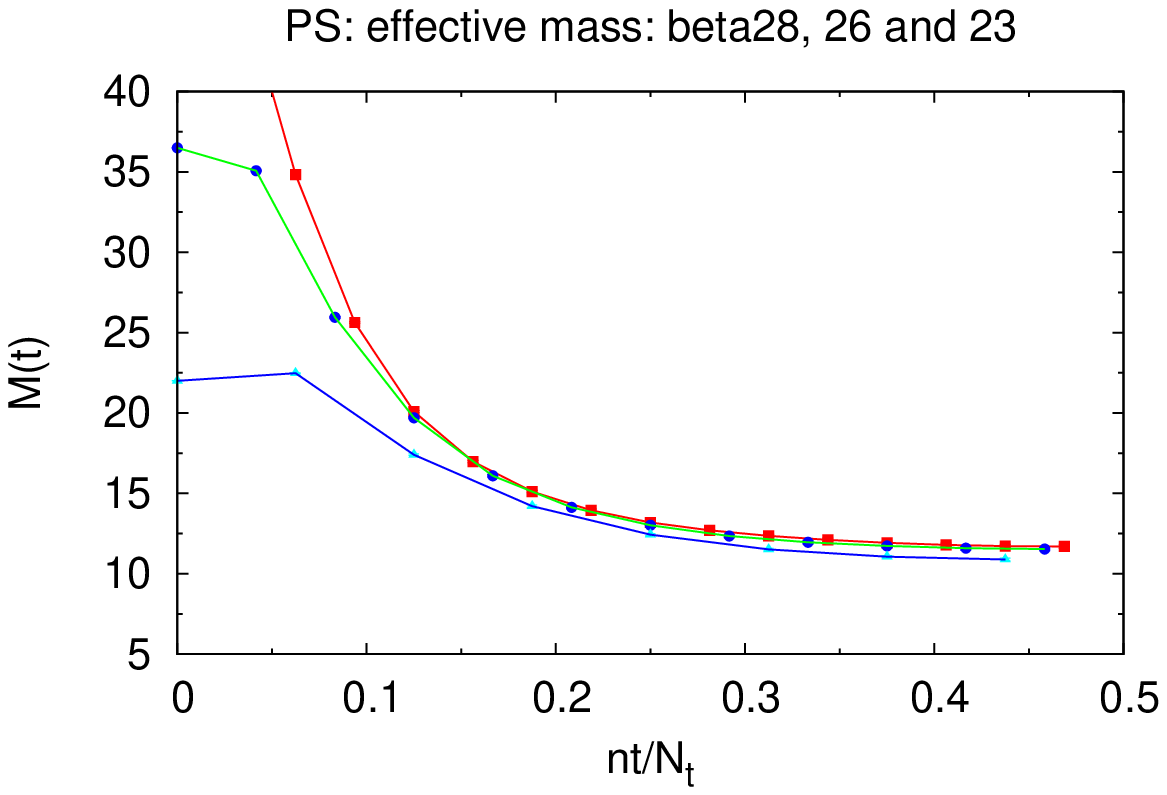}
\hspace{1cm}
\includegraphics[width=7.5cm]{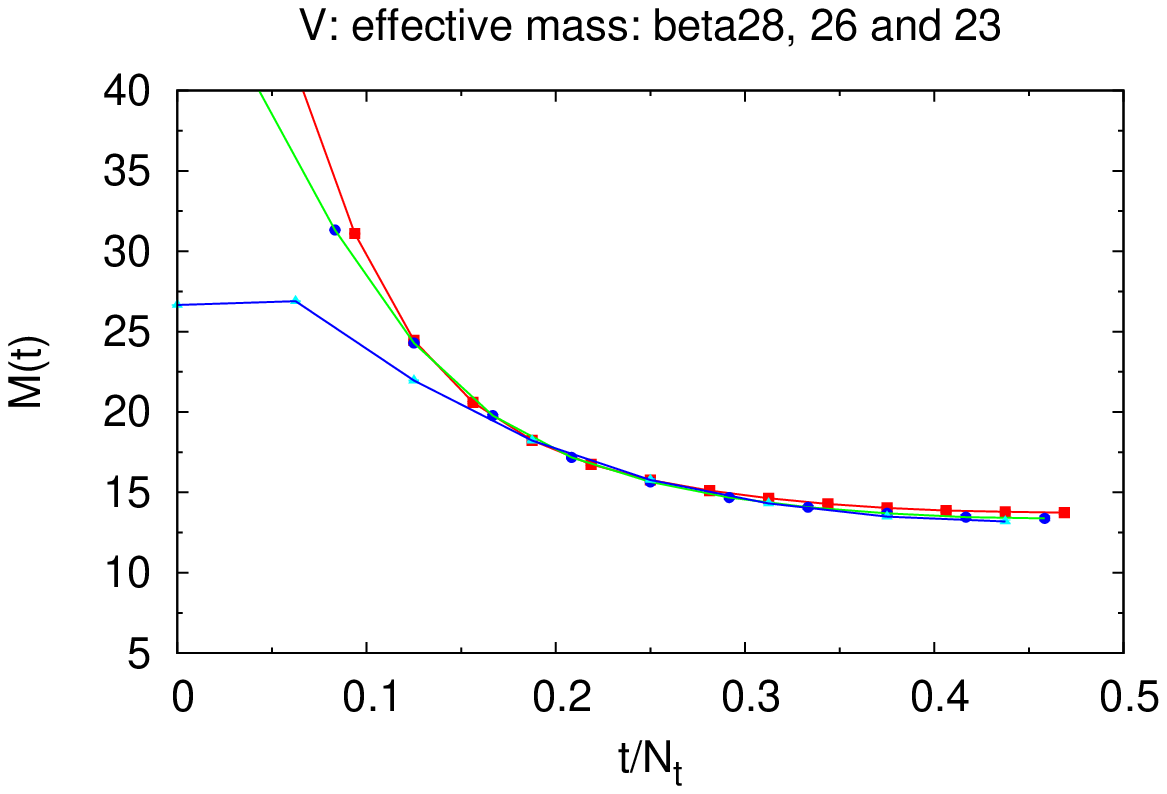}
\caption{(color online)  The effective spatial masses on the $16^3\times 8$ (blue), $24^3\times 12$ (green) and $32^3\times 16$ lattices (red) are overlaid: (left) pseudo-scalar meson; (right) vector meson.
Lines connecting data are for guide of eyes.}  
\label{s-scaling}
\end{figure*}

Now let us show numerical results of the spatial effective masses measured at the critical points in Fig.\ref{s-scaling}.
We plot the scaled spatial effective masses defined in eq.(\ref{scaled effective mass}) in terms of $\tau=n_s/N_s$ 
in order to test the scaling relation
(\ref{RG for s-mass}).
We overlay the data on the three lattices of $32^3\times 16$, $24^3\times 12$ and $16^3\times 8$.

On the left panel we show the effective masses for the pseudo-scalar channel, while on the right panel we show those for the vector channel.
We see that all data are excellently on the scaling curve except for three points at short distance ($n_s =0, 1,2$) on each of the lattices. In particular, the scaling curve for both pseudo-scalar channel and the vector channel on the $32^3\times 16$ and $24^3\times 12$ are within one-standard deviation around $\tau = 0.4$. 
The scaling curve for the vector channel on the $16^3\times 8$ lattice is also almost within one standard deviation, while that for the pseudo-scalar channel on the $16^3\times 8$ lattice is several standard deviations off the other two values. We interpret that the $N=16$ is slightly small to regard it as $N\rightarrow \infty$, in which limit the RG scaling relation holds strictly.

The deviation in short distance behavior is expected to be due to the finite lattice spacing effect. Since it probes the UV behavior, the short distance physics is more sensitive to $1/a$ corrections. We expect that the data at short distance will envelop the limiting scaling curve with increasing $N_t$.

\subsection{temporal effective masses}
\label{temporal}
Next, let us discuss the temporal effective mass $m_t(n_t)$.
The available data points are half of the spatial effective masses.
Even in the case of the largest lattice we take, $32^3\times  16$, the number of the data point is only $8$.
For the other cases they are $6$ and $4$ on $24^3t\times 12$ and $16^3 \times 8$ lattices.
Furthermore since the three points at short distance are affected by the lattice spacing effects, the effective data points we can use are $5$, $3$ and $1$, respectively.
Therefore we do not expect to see a clear scaling.

\begin{figure*}
\includegraphics[width=7.5cm]{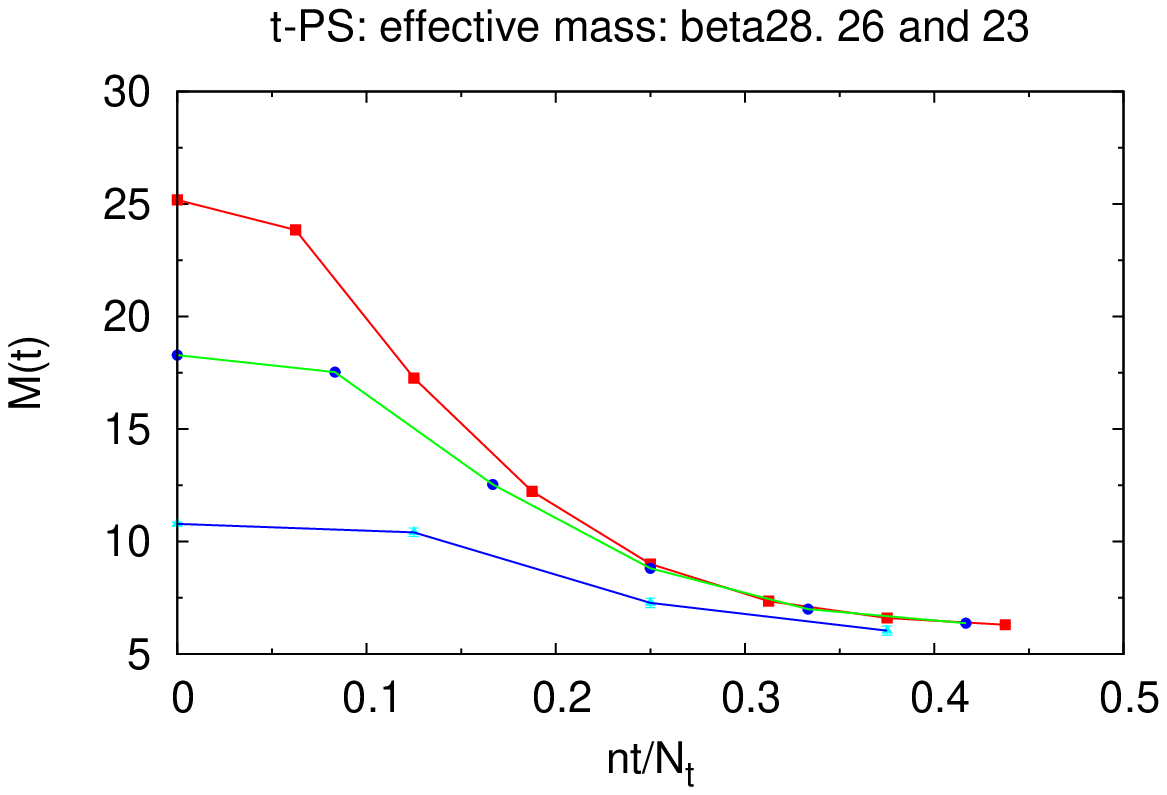}
\hspace{1cm}
\includegraphics[width=7.5cm]{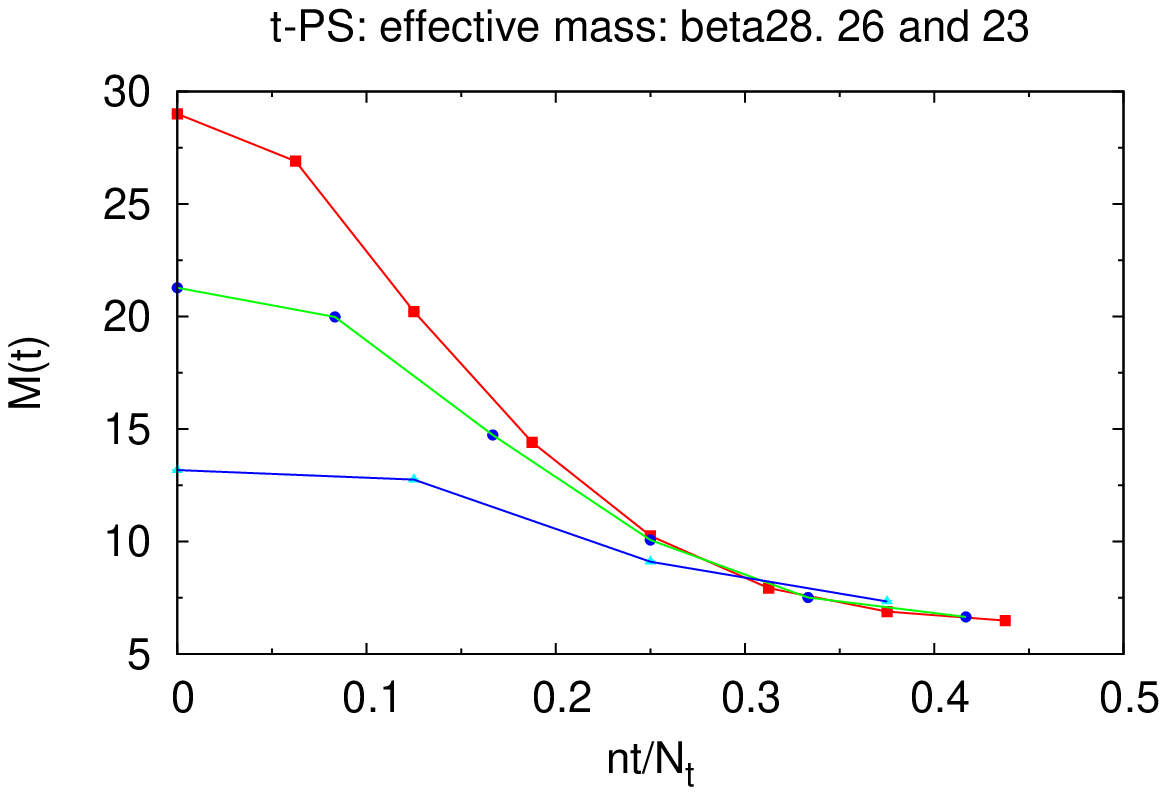}
\caption{(color online)    The effective temporal masses on the $16^3\times 8$ (blue), $24^3\times 12$ (green) and $32^3\times 16$ lattices (red) are overlaid: (left) psuedo-scalar meson; (right) vector meson.
Lines connecting data are for guide of eyes.}
\label{t-scaling}
\end{figure*}

The data are shown in Fig.(\ref{t-scaling}).
On the left panel we present the spatial effective masses of the pseudo-scalar channel, while on the right panel those of the vector channel.
In the pseudo-scalar channel the difference between the $32^3 \times16$ lattice and the $24^3 \times 12$ lattice is order of $1\%,$
which is order of one standard deviation. The difference between the  $16^3\times 8$ lattice and the $32^3\times 16$ lattice is of order $5\%$, which is of order several standard deviations. In the vector channel the differences are of order $2.5\%$.
  Although the clear scaling is yet to be  seen, 
we find the tendency that the larger $N$ curves will converge to the scaling curve. 

\section{Screening mass and power-law corrected Yukawa type decay}

As shown in the preceding section, we find that our numerical results on the three  sizes of lattices are excellently on the scaling curves.
In particular the data on the $32^3 \times 16$ lattice well represents the universal curve $\mathfrak{F_{s}}(\tau)$.
Note the spatial scaling curve corresponds to the screening mass.
It is time to ask the physical meaning of the shape of the scaling curve $\mathfrak{F_{s}}(\tau)$. 

The renormalization group equation itself does not determine the shape of the curve, so it should contain the more detailed physics of the chiral phase transition. Fig.1 shows that the effective masses on the $32^3\times 16$ lattice is monotonously decreasing and does not show a plateau up to the maximum point $n_s=15$ (with the periodic boundary condition). The asymptotic behavior of propagator $G(n_s)$  in the limit $n_s \rightarrow 15$ is well represented by
\begin{equation}
G(n_s)=c \exp(-\hat{m} n_s)/n_s^{\alpha}\label{yukawa_discrete}
\end{equation}

\begin{figure*}
\includegraphics[width=7.5cm]{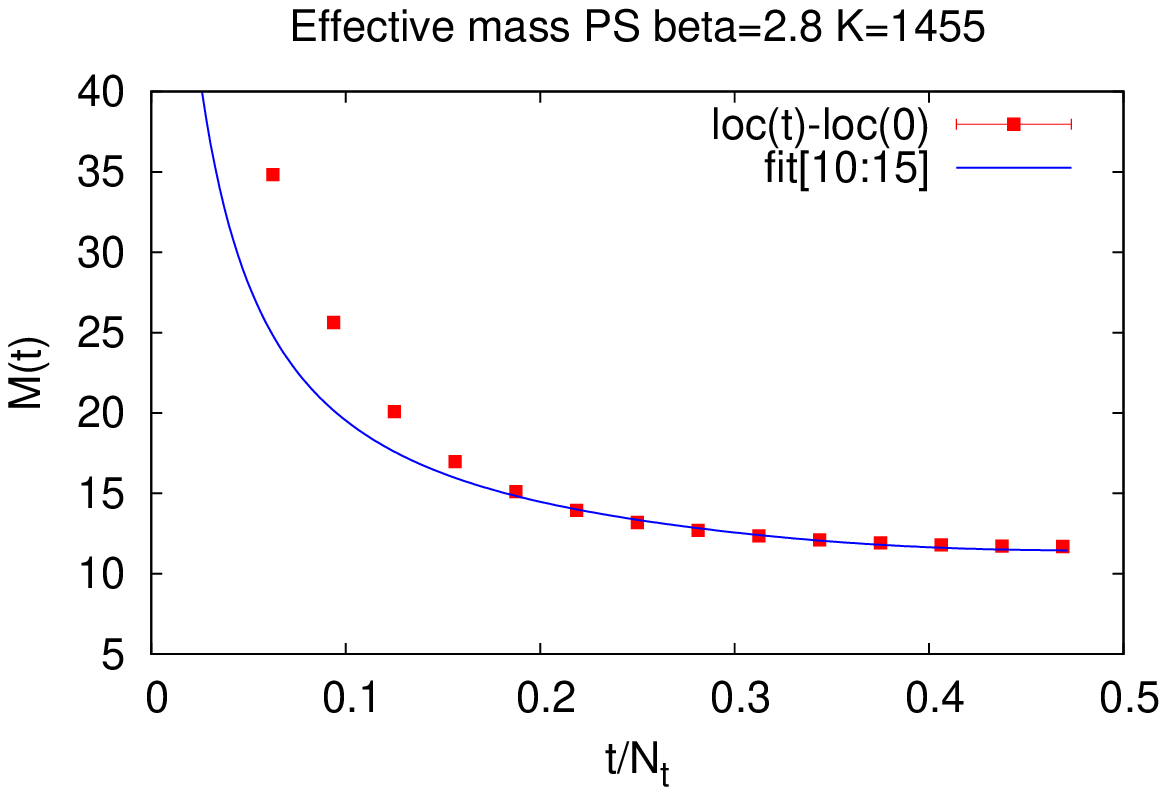}
\hspace{1cm}
\includegraphics[width=7.5cm]{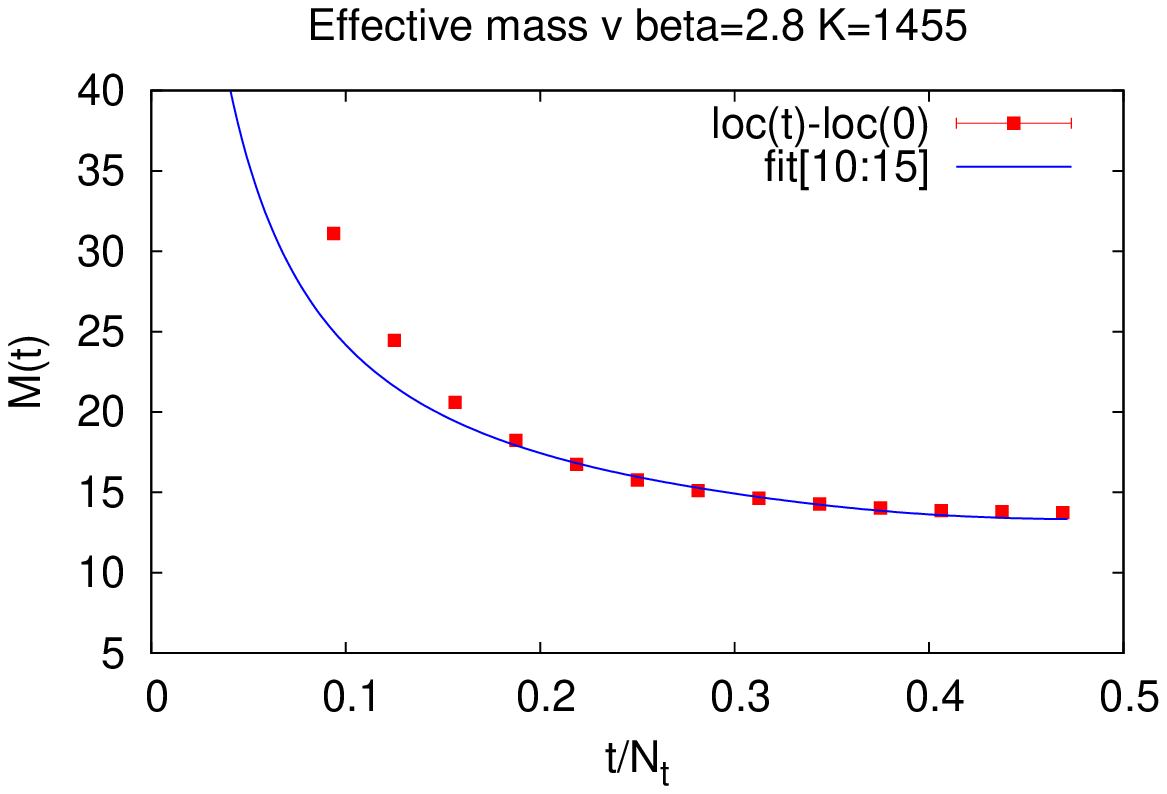}
\caption{(color online)    The effective spatial mass on a  $32^3\times 16$ lattice (red)
and the Yukawa type decaying function (blue) fitted to the data for $10 \le n_s \le 15$; (left) pseudo-scalar meson; (right) vector meson
The fits are $\hat{m}=0.314(6)$ and $\alpha=0.80(8)$ for pseudo-scalar meson and $\hat{m}=0.351(11)$ and $\alpha=1.11(11)$ for vector meson.
They are shown in terms of  $\tau=n_s/N_s$  (x-axis) and $32 \times m_s$ (y-axis).}
\label{yukawa}
\end{figure*}

We fit the data (from $n_s=10$ to $15$) on the $32^3 \times 16$ lattice to the power-law corrected Yukawa type form.
The results of the fit are shown with the data in Fig.~\ref{yukawa} for the pseudo-scalar channel on the left panel and the vector channel on the right panel, respectively.
The best fitting parameters are $\hat{m}=0.314(6)$ and $\alpha=0.80(8)$ for the pseudo-scalar channel and $\hat{m}=0.351(11)$ and $\alpha=1.11(11)$ for the vector channel.
We present them in terms of  $\tau=n_s/N_s$  (x-axis) and $N_s \times m_s$ (y-axis). 

If we fitted the data with a constant assuming a plateau at $n_s=14$ and $15$,
we would get $m=0.366(1)$ and $m=0.429(14)$ for the pseudo-scalar channel and the vector channel respectively. These are about $15-20\%$ larger than the one estimated from the power-law corrected Yukawa type form.
 
In previous articles ~\cite{Ishikawa:2013wf} - \cite{Ishikawa:2013tua}, we pointed out that the propagators show the power-law corrected Yukawa type decay at zero temperature in the $N_f=7, 8, 12$ and $16$ cases and at finite temperature $N_f=2$ case on a $16^3 \times 64$ lattice.
There, we argued that the power-corrections to the exponential decay may be originated from the unparticle nature of the (conformal) fixed point in agreement with the holographic analysis. Based on this theoretical picture, we claimed that  the power-law corrected Yukawa-type decay observed not only for the special aspect ratio $r=1/4$, but for any aspect ratio. Our new results support this claim.

\section{Critical phase transition temperature}
Before we conclude, let us estimate the chiral phase transition temperature $T^{*}$  given by $1/(a N_t )$.
In order to fix the scale, a standard way we also apply is extrapolating the $\rho$ meson mass (in lattice units) to the chiral limit  and comparing it with the experimental value. The spatial lattice size should be the same as that of the finite temperature for a given $\beta$ in a such way that  the Hamiltonian in the  three dimensional space is identical. 
While the lattice size $N_t$ should be large such that $r=N_s/N_t = 1/4.$ 

Due to the CPU time constraint, we focus on the case with the $16^3 \times 64$ lattice at $\beta=2.3$. The gauge coupling constant $\beta$ here is fixed by the chiral phase transition point for the $16^3 \times 8$ lattice that we determined in section 6.

\begin{figure}
\includegraphics[width=7.5cm]{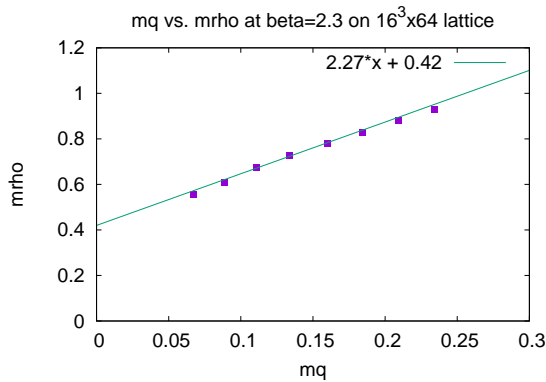}
\hspace{1cm}
\caption{(color online)    The mass of vector meson vs. $m_q$ at $\beta=2.3.$
The fit with the fit range from $K=0,145$ to $K=0.152$.}
\label{beta23}
\end{figure}

The simulations are done at nine hopping parameters from $K=0.140$ to $K=0.152$ (from $m_q=0.372$ to $m_q=0.067$)
listed in Table~\ref{16x64 spectroscopy}.
We fit the rho meson mass  in the linear way $m_\rho = a m_q+c$ to obtain  the vector meson mass in the chiral limit (more precisely; the physical point).
The fit depends on the choice of the fit range of the quark mass. For example, when the fit range is from $K=0.145$ to $K=0.152$,
the rho meson mass extrapolated  takes $m_\rho = 0.42(1)$ with $\chi^2/n f =4.4$.
The fit with the data are shown in Fig.\ref{beta23}.
If we restrict the fit with the condition $\chi^2 \leq 6.0$, $m_\rho=0.400 \sim 0.445\, \,  1/a.$
Using the experimental value  $m_\rho=770 $ MeV as an input, we obtain $1/a=1.73 \sim 1.83$ GeV at $ \beta=2.3.$ This implies that
$a= 0.104 \sim 0.115$ fm and the lattice size in spatial direction is $L_s =1.66 \sim 1.85$ fm.
The transition temperature is estimated  to be $T^ {*}\simeq 215 \sim 240$ MeV.

This value is larger than the value $T_c \sim 170 - 190$ MeV  quoted in \cite{Karsch:1999vy}. See also the more recent review \cite{Kanaya:2010qd} and reference therein. 
However it is notoriously difficult to obtain the common value for the chiral phase transition temperature. 
It is important to perform a similar analysis on larger lattices and take the continuum limit carefully.

\section{summary and discussion}
\label{summary}
We have investigated the nature of the chiral phase transition in the degenerate two flavor QCD using the RG improved gauge action and the Wilson quark action on the $32^3\times 16$, $24^3\times 12$, and $16^3\times 8$ lattices.

We derived the  RG scaling relations at the chiral phase transition point based on Wilson's idea by assuming that the chiral phase transition is second order
\begin{equation}\mathfrak{\check{m}}_s(\tau, g(N), N)=\mathfrak{\check{m}}_s(\tau, g(N{'}),  N{'}).\end{equation}

Our numerical results show that
three effective spatial masses for the pseudo scalar mesons and vector mesons excellently scale.
While the number of the data points are small in the temporal masses to see a clear scaling behavior, the data are consistent with the scaling.
The RG scaling relations are derived with the assumption that the chiral phase transition is second order, and there is no reason to hold if the chiral phase transition is the first order. Our results therefore support the claim that the chiral phase transition is the second order.
Nevertheless, 
we are not able to exclude a possibility that accidentally the  correlation length is very large in a first order transition.

The RG approach is very powerful in investigating the properties of the critical points. It is our future mission to verify the RG scaling relations on larger lattices  such as $48^3\times 24$ to see if the scaling relation persists in agreement with that the chiral phase transition is second order.
In addition, it is important to derive critical exponents and compare them with the theoretically predicted ones to conclude the issue of the order of the chiral phase transition. In particular, the precise knowledge on the critical exponents will reveal whether the $U(1)_A$ symmetry is restored at the chiral phase transition point or not because the (non-)existence of the $U(1)_A$ symmetry affects the critical exponents.

 We would also like to thank A. Ukawa for reading through the article and K. Kanaya for useful discussion.
The calculations were performed on Hitachi SR16000 at KEK under its Large-Scale Simulation Program and HA-PACS computer at CCS, University of Tsukuba under HA-PACS Project for advanced interdisciplinary computational sciences by exa-scale computing technology.

\clearpage

\onecolumngrid

\renewcommand{\topfraction}{1.0}
\renewcommand{\bottomfraction}{1.0}
\renewcommand{\dbltopfraction}{1.0}
\renewcommand{\textfraction}{0.0}
\renewcommand{\floatpagefraction}{1.0}
\renewcommand{\dblfloatpagefraction}{1.0}
\setcounter{topnumber}{8}
\setcounter{bottomnumber}{8}
\setcounter{totalnumber}{8}

\begin{table}
\caption{Simulation parameters at massless quark points $K_c$  on the $32^3\times 16$,  $24^3\times 12$ and $16^3\times 8$ lattices and numerical results for $m_q$, $m_{PS}$, $m_V$ and the chiral condensate $\langle\bar{ \psi}\psi \rangle$. ''traj.'' represents 
the number of trajectories (including thermalization)  and ''acc.''represents the acceptance ratio of the HMC Metropolis test.
The ''NAN'' represents that the simulation cannot be performed due to the zero eigenvalues. See the main text in section IV.}
\begin{tabular}{lcccccccc}
\hline
size & $\beta$ & $K$ & traj. &acc.& $m_q$ & $m_{PS}$ & $m_V$ &$\langle\bar{ \psi}\psi \rangle$ \\
\hline
16x32 & 3.0&   .1435 & 2000&0.84(5)&.0087(2) &  .374(2)  & .408(3) &   0.0087(15)  \\
16x32 &2.9 & .1445 &2000&0.76(4)&   .0021(3) & .379(3) &  .427(7) & 0.0619(7)   \\
16x32 & 2.8 &.1455 & 3000& 0.69(2)&.0052(3)& .365(3)&  .429(3) & 0.0169(14)    \\
16x32&2.7 & NAN\\
\hline
12x24 &2.7 &    .1475 & 2000&0.87(3)&  -.015(1)  &.494(10) &.575(2)   &   -0.060(1)  \\
12x24 &2.6  &   .148 & 3000&0.83(1)& .0091(4) & .481(50 &.668(6)      &  0.032(10)     \\
12x24 &2.5 &   NAN\\
\hline
8x16 & 2.5 & .150 &2500&0.94(1)& .003(1) & .735(11) & .821(15) &  -0.0033(44)\\
8x16 & 2.4 & .152 &3000&0.94(2)& -.001(1) & .718(6) & .822(8) &  -0.0033(44)\\
8x16 & 2.3  &.1547&6000& 0.88(1)&-.009(6)& .680(5)&  .824(6) &  -0.0393(20) \\  
8x16 & 2.2 &  NAN\\
\hline
\label{job parameter}
\end{tabular}
\end{table}

\begin{table}
\caption{ Numerical results for $m_q$, $m_{PS}$, and  $m_V$ on the $16^3\times 64$ lattice at $\beta=2.3$ for determination of the lattice spacing} 
\begin{tabular}{lccc}
\hline
$K$ &  $m_q$ & $m_{PS}$ & $m_V$\\
\hline
.140  &  .372(1)   &  1.113(1) &     1.148(1)\\ 
.145  &  .234(1)   &  0.879(1) &    0.931(2)\\ 
.146  &  .209(1)   &  0.826(1) &   0.879(2)\\
.147  &  .184(1)  &   0.769(1) &    0.828(2)\\  
.148  &  .160(1)  &   0.716(1) &    0.781(1)\\  
.149  &  .134(1)  &   0.656(2) &     0.729(4)\\   
.150  &  .111(1)  &    0.591(1)  &    0.676(3)\\   
.151  &  .089(1)  &   0.523 (2)  &   0.609(5)\\  
.152  &  .067(1)  &   0.448(3)  &  0.556(12)\\  
\hline
\label{16x64 spectroscopy}
\end{tabular}
\end{table}

\end{document}